\documentclass[a4paper,11pt]{article}
\pdfoutput=1 % if your are submitting a pdflatex (i.e. if you have
\usepackage{jcappub} % for details on the use of the package, please
\usepackage[T1]{fontenc} % if needed
\usepackage{comment}

\title{Vacuum Energy Density Measured from Cosmological Data}

\author[a, b, 1]{J.~Prat,} \note{Corresponding author.}
\author[a, b, c, 2]{C. Hogan,} \note{Corresponding author.}
\author[a, b]{C.~Chang,} 
\author[a, b, c]{J.~Frieman}

\affiliation[a]{Department of Astronomy and Astrophysics, University of Chicago, Chicago, IL 60637, USA}
\affiliation[b]{Kavli Institute for Cosmological Physics, University of Chicago, Chicago, IL 60637, USA}
\affiliation[c]{Fermi National Accelerator Laboratory, Batavia, IL 60510, USA}

\emailAdd{jprat@uchicago.edu}
\emailAdd{craighogan@uchicago.edu}
\emailAdd{chihway@kicp.uchicago.edu}
\emailAdd{frieman@fnal.gov}

\abstract{Within the $\Lambda$CDM cosmological model, the absolute value of Einstein's cosmological constant $\Lambda$, sometimes expressed as the gravitating mass-energy density $\rho_\Lambda$ of the physical vacuum, is a fundamental constant of nature, whose accurate measurement  plays a central role in testing some proposed theories of quantum gravity. Several combinations of currently public cosmological data and an assumed flat $\Lambda$CDM cosmological model are used here to make a joint Bayesian inference on the combination of conventional parameters $\Omega_\Lambda h^2$ that corresponds to the absolute physical density  $\rho_\Lambda$. In physical units, we obtain   $\rho_\Lambda = \left(60.3\pm{1.3}\right)\times 10^{-31}{\rm g/cm^3}$, the most accurate constraint to date, with an absolute calibration of cosmological measurements based on CMB temperature. Significantly different values are obtained with calibrations that use a local distance scale, mainly connected to systematic differences in the value of the Hubble constant. It is suggested that future comprehensive cosmological parameter studies assuming the $\Lambda$CDM model include constraints on the vacuum density.}

\begin{document}
\maketitle
\flushbottom

\section{Introduction}
\label{sec:intro}

Most current cosmological measurements are consistent with a dark energy component with equation of state  $p=-\rho c^2$. If that continues to hold with future even more constraining data it would imply that the gravitational effect of dark energy is physically equivalent to Einstein's cosmological constant, $\Lambda$ \cite{Planck2018cosmo, y3-3x2pt}. In this case, the value of $\Lambda$, expressed as the active gravitational mass-energy density  of the physical vacuum, $\rho_\Lambda \equiv \Lambda/8\pi G$, is an absolute constant of nature that can in principle be measured from cosmological data. 

Many widely-studied  multiverse cosmologies posit that $\rho_\Lambda$ is set by random selection, in which case its connection to other physical constants will never be calculable \citep{Weinberg1989,Weinberg2008}.  In this situation, its absolute value is of no particular interest. Indeed, it has not been explicitly addressed in any recent experimental studies. However, it is also possible that the value of $\rho_\Lambda$ could have a precisely defined theoretical relationship to other fundamental physical quantities that can be measured in the laboratory, such as  masses of elementary particles \citep{Hogan2000,bjorken2001,uzan2011}. In this case, we can hope that a successful theory of dark energy could one day resemble other  successful theories\footnote{The most successful  theory, QED, achieves a fractional accuracy of parts per billion in predictions for the anomalous magnetic moment of the electron, compared with other measurements of fine structure constant \citep{PDG2020}.}: it will make predictions for one precisely measured physical quantity, in this case $\rho_\Lambda$, in terms of others, such as Newton's constant $G$ and  the mass of the pion (say), or the mass of the Higgs particle.

In this work we present the most constraining modern cosmological measurement of $\rho_\Lambda$ with an accurately measured covariance, which takes into account the correlations between all the $\Lambda$CDM parameters. We combine different cosmological datasets to estimate values and uncertainties for the actual physical density, including measurements from the Cosmic Microwave Background (CMB), Type Ia Supernovae, Baryonic Acoustic Oscillations (BAO), weak gravitational lensing (WL) and large scale structure (LSS).

The physical density of the vacuum can be expressed as $\rho_\Lambda=\rho_{cr} \Omega_\Lambda = 1.88\times 10^{-29} \Omega_\Lambda h^2$ g/cm$^3$, where the critical density  $\rho_{cr} \equiv 3H_0^2/ 8\pi G $, $\Omega_\Lambda \equiv \rho_\Lambda/\rho_{cr}$, and the scaled Hubble parameter $h\equiv H_0/ (100 {\rm km \ s^{-1} \ Mpc^{-1} })$. Thus, to determine $\rho_\Lambda$, we use cosmic measurements to infer the parameter combination $\Omega_\Lambda h^2$.
%Thus, $\rho_\Lambda = (\Omega_\Lambda h^2) (3/8\pi G)(100 {\rm km \ s^{-1} \ Mpc^{-1} })^2$. Convert this to g/cm cubed.
Previous studies of cosmological parameters have generally adopted  conventional parameter sets that fit for $\Omega_\Lambda$ (or $\Omega_m$) and $h$ separately, except in \cite{Planck2018cosmo} where it had also been fitted before using only \textit{Planck} data. We suggest that estimates of $\rho_\Lambda$ should be integrated into future parameter studies that assume the $\Lambda$CDM model, so that the accuracy of measurements with different data sets can be compared with each other and the results compared with proposed physical theories.

Apart from our choice of parameters,  the analysis here largely follows standard cosmological recipes.  For the current analysis, we adopt the flat $\Lambda$CDM cosmological model, with matter density parameter $\Omega_m=1-\Omega_\Lambda$.
The assumption of flatness, or near-zero mean global curvature, is consistent with the datasets we use, and is a generic prediction of cosmic inflation.

Since $\rho_\Lambda$ depends on a combination of $H_0$ and $\Omega_\Lambda$, its estimated value will be affected by the current tension in the expansion rate parameter. In this work we explore for the first time this dependency and present results for $\rho_\Lambda$ using constraints on $H_0$ coming from the high-redshift Universe, in particular from \textit{Planck}, and compare them with results using local measurements of $H_0$ that rely on the distance ladder from either Cepheid stars calibration \cite{Riess21} or the Tip of the Red Giant Branch calibration \cite{Freedman2021}. 

%Although the analysis here is  a straightforward application of standard techniques, the main motivation represents a conceptual shift,  changing the focus of measurements from  the dark energy equation of state parameter, $w$, a differential measurement,  to $\rho_\Lambda$, an absolute measurement. Viewed in this light, an accurate absolute value of the Hubble constant is not merely an astronomical ``holy grail'' but has a fundamental physical significance, because it determines the density of the  vacuum, whose mean gravity affects the cosmological expansion.

This paper is structured as follows. In Section~\ref{sec:theory} we present the theoretical framework and describe the methodology that we employ to measure $\rho_\Lambda$. In Section~\ref{sec:data} we present the different data sets and likelihoods that we combine to obtain the $\rho_\Lambda$ results and discuss how they are affected by the current $H_0$ tension. We conclude in Section~\ref{sec:summary}.

%\citep{Zeldovich1968,Hogan2000,Perez2018,Perez2019,Hogan2020}.

\section{Methodology}
\label{sec:theory}

We are interested in measuring the dark energy density assuming a flat $\Lambda$CDM Universe, in which case we can express the first Friedmann equation as:
\begin{equation}
    H^2(z) = H^2_0 \left[ \Omega_{m} (1+z)^3 +  (1-\Omega_m) \right],
\end{equation}
where $\Omega_{m} = \rho_m/\rho_{\text{cr}} = \Omega_{\text{CDM}} + \Omega_{b} + \Omega_{\nu}$ is the total matter density today including cold dark matter, baryonic matter and massive neutrinos. In this case, the physical vacuum energy density is given by:
\begin{equation}\label{eq:rho_lambda}
\rho_\Lambda = \frac{3(1-\Omega_{m}) H_0^2}{8\pi G},
\end{equation}
which can also be expressed as $\rho_\Lambda = 1.88\times 10^{-29} (1-\Omega_m) h^2$ g/cm$^3$.
%Each of the dimensionless component parameters is related to physical density via the critical density of the Universe today $\rho_\text{cr}$:
%\begin{equation}
%    \Omega_{\Lambda} = \frac{\rho_{\Lambda}}{\rho_{\text{cr}}},  \quad \Omega_{m} = \frac{\rho_m}{\rho_{\text{cr}}}
%\end{equation}
%with
%\begin{equation}
%   \rho_{\text{cr}} = \frac{3H_0^2}{8\pi G}
%\end{equation}
%Then, the physical value of the dark energy density in a flat $\Lambda$CDM Universe can be expressed as:
%\begin{equation}\label{eq:rho_lambda}
%   \rho_{\Lambda} = \frac{3\Omega_{\Lambda} H_0^2}{8\pi G}.
%\end{equation}
%In a flat Universe $\Omega_{m} + \Omega_\Lambda = 1$. Therefore, we can obtain the dark energy density from Eq.~(\ref{eq:rho_lambda}) knowing the expansion rate and the density parameter of matter today, by substituting $\Omega_{\Lambda} = 1- \Omega_{m}.$ 

To obtain constraints on the $\Lambda$CDM parameters and in particular on the dark energy density we perform a Bayesian analysis. Given some data measurements $\boldsymbol{D}$ for which we have a model $\text{M}$ that depends on a set of parameters $\boldsymbol{p}$ and produces a set of corresponding theoretical predictions $\boldsymbol{T}_\text{M}(\boldsymbol{p})$  we can construct the following Gaussian likelihood:
\begin{equation}
    \mathcal{L} (\boldsymbol{D}| \boldsymbol{p}, \text{M}) \propto e^{-\frac{1}{2} \left[  \left(\boldsymbol{D} - \boldsymbol{T}_\text{M}(\boldsymbol{p}) \right)^\text{T} \text{C}^{-1}    \left(\boldsymbol{D} - \boldsymbol{T}_\text{M}(\boldsymbol{p}) \right)   \right] },
\end{equation}
where $\text{C}$ is the covariance between the measurements. Using Bayes theorem,  we can obtain the posterior probability distribution for the parameters $\boldsymbol{p}$ from the likelihood:
\begin{equation}
    P(\boldsymbol{p}| \boldsymbol{D}, \text{M})  \propto      \mathcal{L} (\boldsymbol{D}| \boldsymbol{p}, \text{M})  P(\boldsymbol{p}|  \text{M}), 
\end{equation}
where $P(\boldsymbol{p}|  \text{M}) $ is a prior probability distribution on the parameters  $\boldsymbol{p}$. In this work we use measurements $\boldsymbol{D}$ from different probes including  type Ia Supernovae, Baryonic Acoustic Oscillations (BAO), Cosmic Microwave Background anisotropies  (CMB), large-scale structure (LSS) and weak lensing (WL). In each of the cases we construct a different likelihood which includes a different set of parameters $\boldsymbol{p}$ to model the data. In all cases  $\boldsymbol{p}$ will include the $\Lambda$CDM parameters, assuming flatness. Specifically, we sample over the cosmological parameters shown in Table.~\ref{tab:priors_lcdm}. We assume 3 species of massive neutrinos.  We also sample over the reionization optical depth $\tau$. In the following section we will detail each of the measurements we use in this paper, and in Table~\ref{tab:priors_nuisance} we list the priors for the extra nuisance parameters we assume in each case. 

We use the public version of \textsc{CosmoSIS} in the \texttt{develop} branch\footnote{\texttt{https://bitbucket.org/joezuntz/cosmosis-standard-library/src/develop/}} \citep{Zuntz:2014csq} to sample the different likelihoods. We use the \textsc{Polychord} sampler \citep{Polychord1,Polychord2} with the following settings \texttt{feedback = 3}, \texttt{fast\_fraction = 0.1}, \texttt{live\_points = 500}, \texttt{num\_repeats=60}, \texttt{tolerance=0.1}, which were validated in \citep{Lemos2022}. We use \textsc{ChainConsumer} \citep{Hinton2016} to visualize the  Markov chain Monte Carlo (MCMC) and to compute the mean of the posteriors.

\begin{table}
    \centering
    \begin{tabular}{cl}
        \hline
		Parameter & Prior \\ 
		\hline
		 $\Omega_m$ & Flat [0.1, 0.9]  \\ 
		 $h$   & Flat [0.55, 0.91] \\
		 $\Omega_b$ & Flat [0.03, 0.07] \\
		 $n_s$ & Flat [0.87,  1.07] \\
		 $A_s$ & Flat [0.5$\times 10^{-9}$, 5$\times 10^{-9}$] \\
		 $\Omega_\nu h^2$  & Flat [0.0006, 0.00644]\\
		 $\tau$  & Flat [0.01, 0.8] \\
		\hline
    \end{tabular}
     \caption{$\Lambda$CDM parameter priors used in this work. \label{tab:priors_lcdm}}
\end{table}

\section{Data and results}
\label{sec:data}

\begin{figure*}
\begin{center}
\includegraphics[width=1.\textwidth]{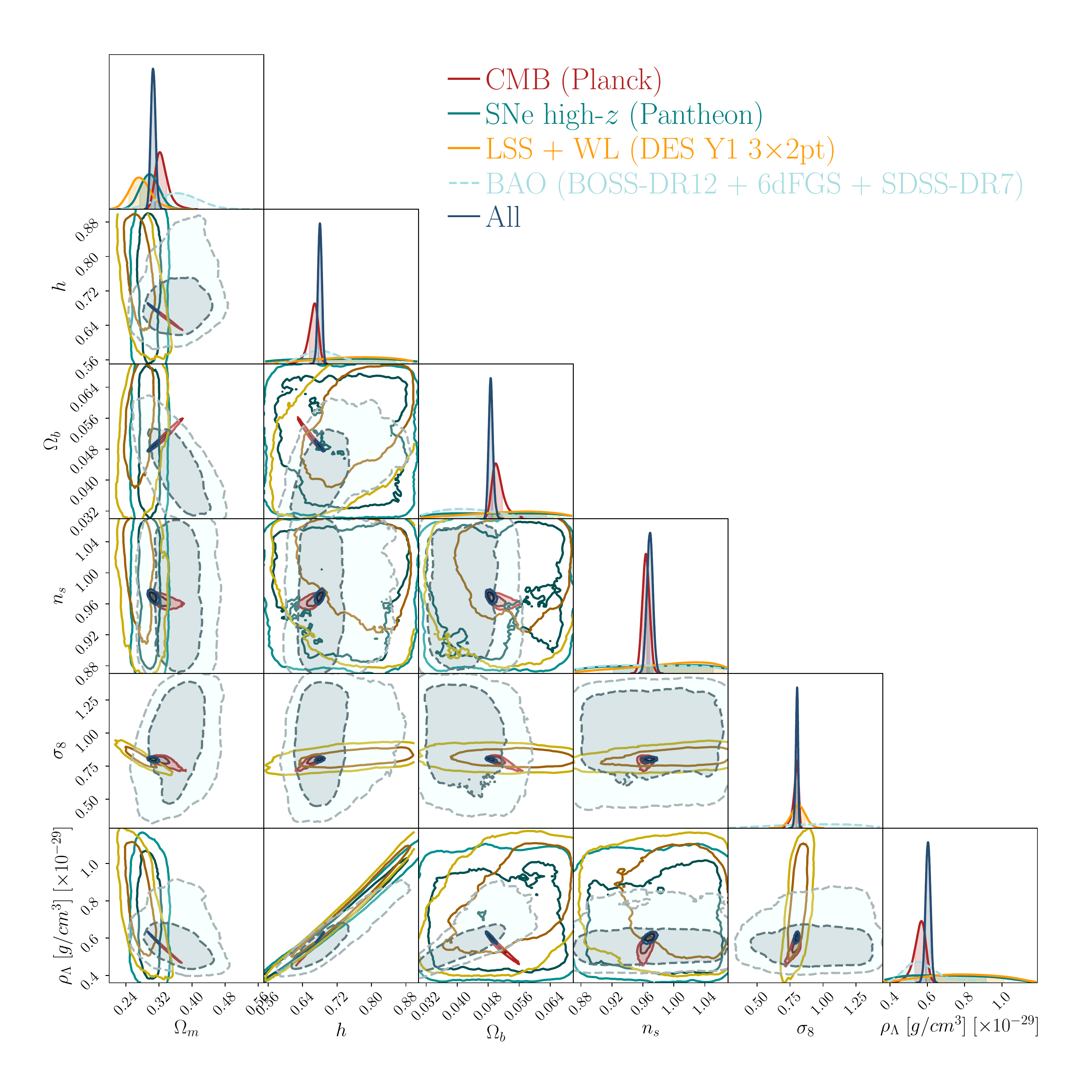}
\caption{Posterior parameter distributions, showing the correlations between the different $\Lambda$CDM parameters using each of the probes and their combination (labelled ``All''). $\rho_\Lambda$ is the mass density of the cosmological constant expressed in physical units, which is proportional to (1-$\Omega_m) h^2$, see Eq.~(\ref{eq:rho_lambda}). The contours are posteriors from Bayesian likelihood functions. The fiducial result includes: (i) the \textit{Planck} likelihood \citep{Planck2018cosmo} including lensing, (ii) the Pantheon type Ia supernovae distance measurements between $0.01 < z < 2.3$ \citep{Scolnic:2017caz}, (iii) the large-scale structure (LSS) and Weak Lensing (WL) measurements from the Dark Energy Survey Y1 data \citep{y1-3x2pt} and (iv) the baryonic acoustic oscillation (BAO) measurements from BOSS-DR12, 6dFGS and SDSS-DR7 \citep{Alam2017, Ross2015, Beutler2011}.}
\label{fig:cosmo_allpars_comparison}
\end{center}
\end{figure*}

\begin{table*}
    \centering

    \label{tab:lcdm_all_ext}
    \scalebox{1.}{
    \begin{tabular}{cccccccc}
       
		Parameter &  All & Low-$z$, Local $H_0$ (Cepheids) & Low-$z$, Local $H_0$ (TRGB) \\
		 \hline 
		$\rho_\Lambda$ $[g/cm^3]$  & $\left(60.3\pm 1.3 \right) \times 10^{-31}$ &  $\left( 70.3^{+2.7}_{-3.1} \right) \times 10^{-31}$  & $\left( 63.3^{+3.7}_{-2.9} \right) \times 10^{-31}$ \\
		
		$\Omega_\Lambda h^2$ &  $0.3212^{+0.0070}_{-0.0067}$ &  $0.374^{+0.014}_{-0.017}$ & $0.337^{+0.020}_{-0.015}$ \\
		
		$\Omega_m$  & $ 0.3055^{+0.0060}_{-0.0054}$ & $0.294\pm 0.014$ & $0.294^{+0.016}_{-0.012}$\\
		
		$h$  & $0.6803^{+0.0047}_{-0.0044}$ &  $0.727\pm 0.013$ &  $0.694\pm 0.016$\\
		
		$\Omega_b$  &   $ 0.04865^{+0.00056}_{-0.00055} $  &  $0.0567\pm 0.0036$ &  $ 0.0534^{0.0034}_{-0.0041} $\\
		
		$n_s$  & $ 0.9694^{+0.0036}_{-0.0042}$  & $0.982^{+0.064}_{-0.056}$  & $1.005^{+0.048}_{-0.068}$ \\
		
		$\sigma_8$ & $ 0.8041^{+0.0076}_{-0.0098} $& $0.779^{+0.037}_{-0.031}$    & $0.775^{+0.035}_{-0.032}$ \\ 
		\hline
		\hline
    \end{tabular}}
        \caption{$\Lambda$CDM parameter results. The row labelled ``All'' is our fiducial result and includes  \textit{Planck} 2018 with lensing, Pantheon supernovae, BAO from BOSS-DR12, 6dFGS and SDSS-DR7, and the DES Y1 3$\times$2pt likelihood. The row labelled  ``Low-$z$, local $H_0$ (Cepheids)'' uses the same data as the fiducial results except replacing the \textit{Planck} likelihood with local $H_0$ measurements from \cite{Riess21}, that are based on a Cepheid variable star calibration of the distance ladder. ``Low-$z$, local $H_0$ (TRGB)'' uses the same data as the fiducial results except  replacing the \textit{Planck} likelihood with local $H_0$ measurements from \cite{Freedman2021}, that are based on the Tip of the Red Giant Branch (TRGB) calibration of the distance ladder.}
\end{table*}

We use publicly available data from the Pantheon supernova sample, the Sloan Digital Sky Survey (SDSS), \textit{Planck} and the Dark Energy Survey (DES). We describe below the details of each likelihood:

\begin{itemize}
    \item \textbf{Type Ia Supernovae}: We use relative distance measurements from the Pantheon sample \citep{Scolnic:2017caz}, which comprises 1048 spectroscopically confirmed SNe Ia in the redshift range $0.01 < z < 2.3$. For this likelihood, we sample over the $\Lambda$CDM parameters shown in Table~\ref{tab:priors_lcdm} and an additional nuisance parameter $m$ to marginalize over the uncertainty on the absolute magnitude calibration of the entire sample, as detailed in Table~\ref{tab:priors_nuisance}. This procedure uses the supernovae as a constraint on late-time acceleration, and does not depend on local calibration of their absolute distance, which is discussed in Sec~\ref{sec:h0}.  
    \item \textbf{CMB}: We use the likelihood from the \textit{Planck} 2018 data release \citep{Aghanim:2019ame, Aghanim:2018eyx} with the same specifications as in \cite{y3-3x2pt}. This includes the following likelihoods: (i) the \texttt{Plik} likelihood of the temperature power spectrum, $C_\ell^{TT}$, in the angular multipole range $30 \leq \ell \leq 2508$ and the $E-$mode power spectrum, $C_\ell^{EE}$, and the cross power-spectrum between temperature and the $E$-mode, $C_\ell^{TE}$, in the range $30 \leq \ell \leq 1996$; (ii) the \texttt{Commander} likelihood of the temperature power spectrum, $C_\ell^{TT}$, in the range $2 \leq \ell \leq 29$; (iii) the \texttt{SimAll} likelihood of the $E-$mode power spectrum, $C_\ell^{EE}$, in the range $2 \leq \ell \leq 29$; and (iv) the lensing potential likelihood, $C_\ell^{\phi \phi}$, in the range $8 \leq \ell \leq 400$. Besides sampling over the $\Lambda$CDM paramters from Table~\ref{tab:priors_lcdm}, we also sample over all the usual \text{Planck} nuisance parameters, as shown in Table~\ref{tab:priors_nuisance}. 
    \item \textbf{BAO}: We use the BAO measurements from BOSS Data Release 12 (DR12) \citep{Alam2017}, the SDSS Data Release 7 (DR7) Main Galaxy Sample \citep{Ross2015}, and the 6dF Galaxy Survey \citep{Beutler2011}. This likelihood does not require any additional nuisance parameters. 
    \item \textbf{WL and LSS}: We use the DES Y1 3$\times$2pt likelihood, which includes measurements from weak lensing and large-scale structure. Specifically, we use three two-point correlation functions: (i) the redMaGIC galaxy clustering auto-correlation function, i.e. $w(\theta)$; (ii) the cross-correlation between source galaxy shapes and foreground redMaGIC galaxy positions, i.e., the galaxy-galaxy lensing measurement $\gamma_t (\theta)$; and (iii) the auto-correlation of the shapes of the source galaxy sample, i.e., the cosmic shear measurements $\xi_+$ and $\xi_-$. This likelihood includes additional free nuisance parameters, which are all listed in Table~\ref{tab:priors_nuisance} together with their priors. We use the same model and priors as in  the DES Y1 original 3$\times$2pt analysis \cite{y1-3x2pt} except that here we are assuming that three species of neutrinos are massive, while in the published DES Y1 3$\times$2pt results only one species was assumed to be massive. We check in Appendix~\ref{app:neutrino} that this choice does not significantly affect the DES Y1 3$\times$2pt posterior. \\
\end{itemize}
Note that redshift space distortions (RSD) are not included in this analysis but they also have the potential to constrain $\rho_{\Lambda}$ in future analyses. Our fiducial results are obtained combining all the above probes and are shown in Fig.~\ref{fig:cosmo_allpars_comparison}. We also display the results from each of the probes individually to show the relative importance of each of them and which parameters they constrain the most. The \textit{Planck} posteriors are the tightest in all the $\Lambda$CDM parameters. The Pantheon supernova sample mostly constrains $\Omega_m$. The results from DES Y1 3$\times$2pt are particularly constraining in the $\sigma_8$-$\Omega_m$ plane. The BAO results constrain $\Omega_m$, $\Omega_b$ and $H_0$. 
The numerical results are given in Table~\ref{tab:lcdm_all_ext}. The main result using the combination of all the probes is
\begin{equation}
\Omega_\Lambda h^2 = 0.3212^{+0.0070}_{-0.0067}.
\end{equation}
In physical units this translates to:
\begin{equation}
  \rho_\Lambda = \left(60.3\pm{1.3}\right)\times 10^{-31}{\rm g/cm^3},  
\end{equation}
a 2.2\% measurement, which is the most constraining value reported up to date. The equivalent  physical value of $\Lambda$ itself is then:
\begin{equation}
    \Lambda \equiv 
    8\pi G \rho_\Lambda = 
    \left(1.01 \pm 0.02\right) \times 10^{-35} {\rm s}^{-2}.
\end{equation}
Finally, in distance units we obtain:
\begin{equation}
     \frac{\Lambda}{c^2} = \left(1.13 \pm 0.02\right) \times 10^{-56} {\rm cm}^{-2}.
\end{equation}

\subsection{Systematic calibration uncertainty related to $H_0$ tension} \label{sec:h0}

\begin{figure*}
\begin{center}
\includegraphics[width=0.9\textwidth]{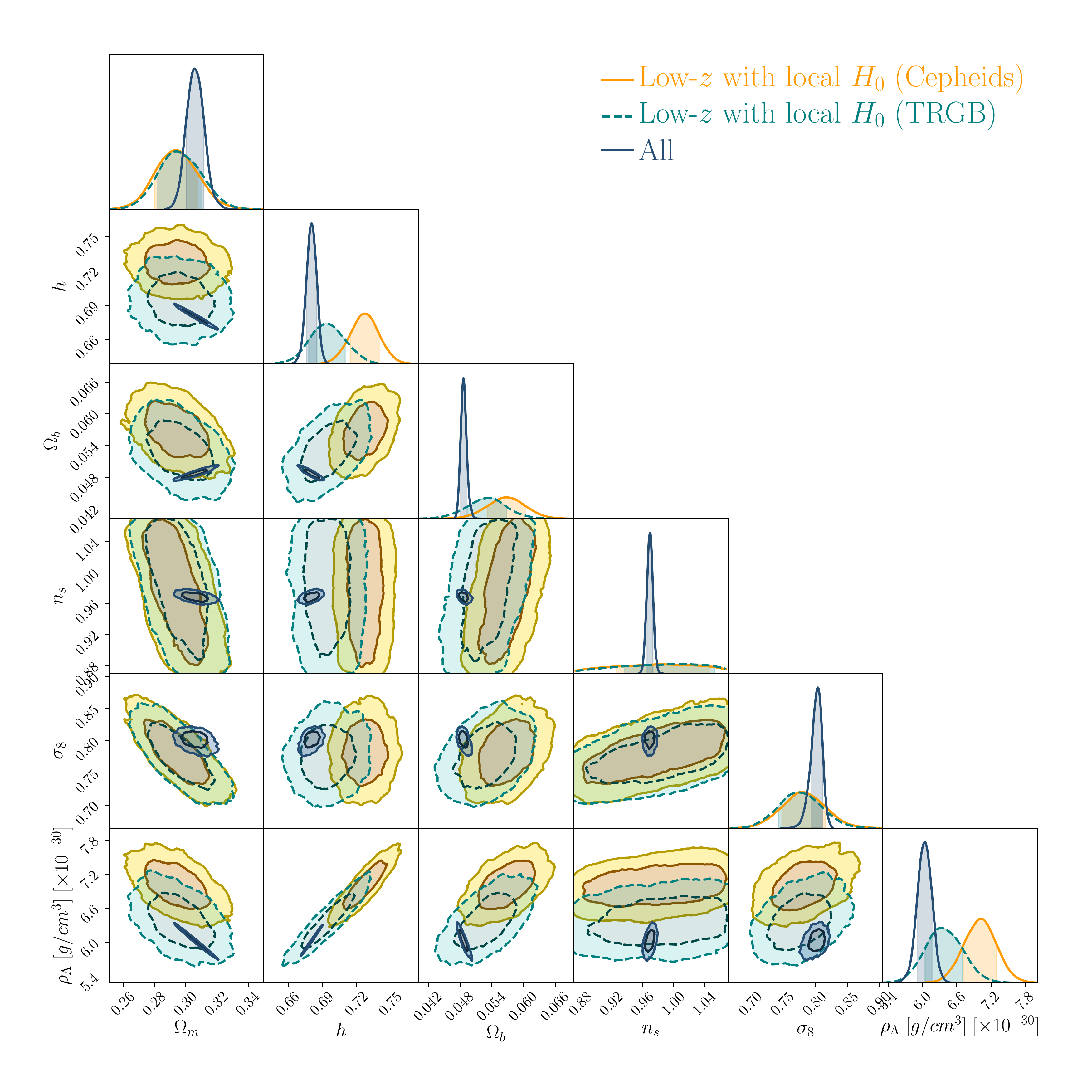}
\caption{This figure shows how the $\rho_\Lambda$ results are affected by the current $H_0$ tension. 
The yellow posterior corresponds to using the same data as the fiducial results, which are displayed in blue, except that we remove the \textit{Planck} likelihood and add instead the local $H_0$ measurements from \cite{Riess21}, which are based on a distance ladder calibration using Cepheid variable stars. The green dashed posterior uses the $H_0$ results published in \cite{Freedman2021} that are based on the Tip of the Red Giant Branch (TRGB) calibration instead. }
\label{fig:cosmo_allpars_H0}
\end{center}
\end{figure*}

Since $\rho_\Lambda$ is obtained from a combination of the parameters $\Omega_m$ and $H_0$, its inferred value is connected to the value of the present-day expansion rate, $H_0$.
Thus, inference of the absolute physical value of $\rho_\Lambda$ inherits uncertainty related to the current tension between different measurements of $H_0$. This uncertainty has a different character from the measurement uncertainty already included in the analysis above.
%The results quoted above are dependent on a measurement of  $H_0$ that is ultimately calibrated by direct local measurement of the present-day CMB temperature. That is, CMB data are fit to a cosmological model,  which yields a value for $H_0$, the expansion rate at the time when the CMB reaches a temperature equal to the observed value. The present-day CMB absolute temperature is measured with high fractional accuracy and contributes negligibly to the total uncertainty. In the context of the minimal flat-$\Lambda$CDM cosmological model,  which is a good fit to the CMB data, the high-$z$-calibrated constraints on $H_0$ and $\rho_\Lambda$ given above are self-consistent. \textit{In this context, the above procedure thus provides a comprehensive estimate of  total uncertainty, including systematic calibration errors. }
%However, this estimate yields values of $H_0$  significantly lower than those obtained from some local measurements based on direct geometrical distances.
Here we compare the fiducial results on $\rho_\Lambda$ presented in Fig.~\ref{fig:cosmo_allpars_comparison} with results obtained with only low-redshift data and local $H_0$ measurements. In particular we use the following likelihoods:
\begin{itemize}
    \item \textbf{Local distance ladder measurements (Cepheids)}: We use the result from \cite{Riess21} of $H_0 = 73.0 \pm 1.4$ km s$^{-1}$ Mpc$^{-1}$, which is obtained using a distance ladder starting from the geometric calibration using GAIA EDR3 parallaxes that are used to calibrate the distance to Cepheids stars which are then used to calibrate the type Ia Supernova absolute magnitude.  
    \item \textbf{Local distance ladder measurements (TRGB)}: We use the result from \cite{Freedman2021} of $H_0 = 69.8 \pm 1.7$ km s$^{-1}$ Mpc$^{-1}$, which is obtained using a distance ladder starting also from the geometric calibration using GAIA EDR3 parallaxes. These are then used to calibrate distances to galaxies that are hosts to type Ia supernovae using the Tip of the Red Giant Branch (TRGB) method.
\end{itemize}

We present the comparison of the fiducial results using high redshift \textit{Planck} data vs. each of the low redshift measurements in Fig.~\ref{fig:cosmo_allpars_H0}. In that figure, the yellow posterior corresponds to using the same data as the fiducial results displayed in blue except for removing the \textit{Planck} likelihood and adding instead the local $H_0$ measurement from \cite{Riess21}. The green posterior is analogous to the yellow one, using instead the $H_0$ value from \cite{Freedman2021}. Note we are able to combine the local $H_0$ measurements with the Pantheon, DES Y1 3$\times$2pt and BAO measurements because they are consistent with each other. This is mainly due to the fact that Pantheon supernovae and the DES Y1 3$\times$2pt posteriors are mostly flat in $H_0$ (see Fig.~\ref{fig:cosmo_allpars_comparison}) and the BAO one is quite broad in $H_0$, with the values of $H_0$ from local distance ladder measurements being within the 1-$\sigma$ region of the posterior. 

The numerical results using the local $H_0$ measurements are shown in the lower two rows of Table \ref{tab:lcdm_all_ext}. Using the local measurements based on the Cepheid calibration yields $\rho_\Lambda = \left( 70.3^{+2.7}_{-3.1} \right) \times 10^{-31}$ g/cm$^3$, significantly higher than the value obtained using the \textit{Planck} likelihood; the two values are discrepant by 3.1 $\sigma$ in the 1-D posterior. Using the analogous data set with the TRGB calibration results in $\rho_\Lambda = \left( 63.3^{+3.7}_{-2.9} \right) \times 10^{-31}$ g/cm$^3$. This is just slightly higher than the fiducial result and is statistically consistent with it. The results using local measurements are approximately two times less precise than the one using the high-$z$ data. Not surprisingly, the combined-probe values of $h$ in the lower two rows of Table \ref{tab:lcdm_all_ext} are pulled only slightly from the Cepheid and TRGB values, since the other low-redshift measurements do not constrain $H_0$ with high precision. However, the combination of the other low-$z$ measurements with the local $H_0$ measurements does require a value of the baryon density, $\Omega_b$, higher than predicted by either the CMB or Big Bang Nucleosynthesis, particularly for the Cepheid measurement.

 Some missing ingredient  is apparently needed to reconcile the CMB 
  values of $H_0$  with  Cepheid-based measurements of  the local distance scale. 
  If the difference is resolved with other calibrations of the local distance scale \citep{Freedman2021}, the value of $\rho_\Lambda$ given above is not changed.  If the difference turns out to  be due to  significant departures  from the flat $\Lambda$CDM model, it would significantly impact the  assumptions we have used to estimate $\rho_\Lambda$; our entire measurement procedure is predicated on the validity of $\Lambda$CDM as a background cosmology.

\subsection{Dimensionless value of $\rho_\Lambda$ in QCD-normalized models}

As an example of the utility of the current exercise --- and an aspirational challenge to theory ---  we can quote a measurement of a pure number that according to some scenarios should be computable from a combination of numerical QCD, normalized to the physical mass of the pion, and a  causally-coherent theory of quantum gravity (see Appendix~\ref{sec:proposals} for a summary on some examples of theoretical proposals that might be developed with sufficient precision for a meaningful test based on an absolute measurement of $\rho_\Lambda$).

Using our fiducial result for $\rho_\Lambda$,  the coefficient $\Lambda_\pi$ from Eq.~(\ref{QCDscaling}) has the value
  \begin{equation}
     \Lambda_\pi = (5.11 \pm 0.11)\times 10^{-4}, 
  \end{equation}
 assuming $139.57061 \pm 0.00024$ MeV/$c^2$ for the mass of a charged pion, from \cite{PDG2020}. The uncertainty of the mass of the pion in this calculation is not significant, and so the uncertainty is dominated by the measurement of $\rho_\Lambda$. If instead we use the $\rho_\Lambda$ result from local $H_0$ measurements from \cite{Riess21}, we obtain $\Lambda_\pi = (5.96 \pm 0.25)\times 10^{-4}$.

\section{Summary and Conclusions}
\label{sec:summary}

$\Lambda$CDM is the current standard cosmological model, which is so far consistent with the most recent and constraining observations, e.g. \cite{y3-3x2pt}. In this model, the energy density of the dark energy component, $\rho_\Lambda$, is an absolute constant of nature which is just a different projection of standard $\Lambda$CDM parameters, $\rho_\Lambda \propto (1-\Omega_m) h^2$. In this paper we report the most constraining measurement to date of $\rho_\Lambda$ in physical units assuming a flat $\Lambda$CDM model.

Using public data sets including CMB observations from Planck, the Pantheon sample of type Ia supernovae, BAO measurements from BOSS Data Release 12, the SDSS Data Release 7 Main Galaxy Sample and the 6dF Galaxy Survey, and large-scale structure and weak lensing measurements from DES Y1, we find a value of $\rho_\Lambda = \left(60.3\pm{1.3}\right)\times 10^{-31}{\rm g/cm^3}$, with an accurately measured covariance that accounts for all the correlations between $\Lambda$CDM parameters. Using local distance ladder measurements of $H_0$ instead of Planck CMB data, we instead find $\rho_\Lambda = (70.3+2.7-3.1) \times 10^{-31}{\rm g/cm^3}$ using Cepheids or $\rho_\Lambda = (63.3+3.7-2.9) \times 10^{-31}{\rm g/cm^3}$ using the Tip of the Red Giant Branch.

We propose that results on this quantity should be included when reporting cosmological results that assume the $\Lambda$CDM model since, in the context of this well-supported model, 
$\rho_\Lambda$ is a fundamental constant of nature. It therefore provides a target for any physical theory that purports to explain the value of the cosmological constant. In the future, estimates of $\rho_\Lambda$ will allow tests of proposed physical theories that aim to connect this constant of nature with others, and in particular to explain the famous $\sim$122-order-of-magnitude mismatch of measured vacuum energy density with a na\"ive estimate for quantum field zero-point energy based on a Planck scale UV cutoff. Comparison with an accurate measurement of $\rho_\Lambda$ will be essential to validate or refute such theories. At present, the precision  of the theories is far behind that of the data (see Appendix~\ref{sec:proposals}). Cosmological measurements of $\rho_\Lambda$ set a standard of precision that provide a quantitative goal for future theoretical development.

\appendix

\begin{table*}
    \centering
    \label{tab:priors_nuisance}
    \begin{tabular}{cl}
        \hline
		Parameter & Prior \\
		\hline 
		\vspace{0.2cm} 
		\textbf{Pantheon SN} \\ 
		 $m$ (Absolute magnitude calibration) & Flat [-20, -18]  \\ 
		\hline  
		\vspace{0.2 cm}  
		\textbf{Planck 2018} \\ 
		
		$a_\text{planck}$  (Total calibration) & Gauss (1.0, 0.0025) \\
		$a_{\text{CIB}_{217}}$ (Spectral index of the CIB) &  Flat [0, 200] \\
		$\xi_{\text{SZ}-{\text{CIB}}}$ (TSZ-CIB template amplitude) & Flat [0, 1] \\
 		$a_\text{SZ}$ (Thermal SZ amplitude at 143 GH) & Flat [0, 10] \\
 		$ps_{a_{100-100}}$ (Point source amplitude at 100 GHz) & Flat [0, 400] \\
 		$ps_{a_{143-143}}$ (Point source amplitude at 143 GHz) & Flat [0, 400] \\
 		$ps_{a_{143-217}}$ (Point source amplitude at 143$\times$217 GHz) & Flat [0, 400] \\
 		$ps_{a_{217-217}}$ (Point source amplitude at 217 GHz) & Flat [0, 400] \\
 		$ksz_\text{norm}$ (Kinetic SZ amplitude at 143 GH) & Flat [0, 10] \\
 		$\text{gal}545_{a_{100}}$ (Dust amplitude at 100 GHz at $\ell=200$ for $EE$) & Gauss (8.6, 2) \\
 		$\text{gal}545_{a_{143}}$ (Dust amplitude at 143 GHz at $\ell=200$ for $EE$) & Gauss (10.6, 2) \\
 		$\text{gal}545_{a_{143-217}}$ (Dust amplitude at 143$\times$217 GHz at $\ell=200$ for $EE$) & Gauss (23.5, 8.5) \\
 		$\text{gal}545_{a_{217}}$ (Dust amplitude at 217 GHz at $\ell=200$ for $EE$) & Gauss (91.9, 20) \\
 		$\text{calib}_{100t}$ (Relative power spectrum calibration factor 100/143) & Gauss (1.0002, 0.0007) \\
 		$\text{calib}_{217t}$ (Relative power spectrum calibration factor 217/143) & Gauss (0.99805, 0.00065) \\
 		$\text{galf}_{TE_{a_{100}}}$ (Dust amplitude at 100 GHz at $\ell=500$ for $TE$) & Gauss (0.13, 0.042) \\
 		$\text{galf}_{TE_{a_{100-143}}}$ (Dust amplitude at 100$\times$143 GHz at $\ell=500$ for $TE$) & Gauss (0.13, 0.036) \\
 		$\text{galf}_{TE_{a_{100-217}}}$  (Dust amplitude at 100$\times$217 GHz at $\ell=500$ for $TE$) & Gauss (0.46, 0.09) \\
 		$\text{galf}_{TE_{a_{143}}}$   (Dust amplitude at 143 GHz at $\ell=500$ for $TE$) & Gauss (0.207, 0.072) \\
 		$\text{galf}_{TE_{a_{143-217}}}$   (Dust amplitude at 143$\times$217 GHz at $\ell=500$ for $TE$) & Gauss (0.69, 0.09) \\
 		$\text{galf}_{TE_{a_{217}}}$   (Dust amplitude at 217 GHz at $\ell=500$ for $TE$) & Gauss (1.938, 0.54) \\
		\hline
		\vspace{0.2 cm}  
		\textbf{DES Y1 WL + LSS (3$\times$ 2pt)} \\ 
		Lens galaxy bias per redshift bin (5 bins), $b_i$ & Flat [0.8, 3] \\
		Intrinsic alignment amplitude (NLA), $A$ & Flat [$- 5$, 5] \\
		Intrinsic alignment $z$-scaling (NLA), $\nu$ & Flat [$- 5$, 5] \\
		Lens photo-$z$ shift, $\Delta z_l^1$ & Gauss (0.008, 0.007) \\
		Lens photo-$z$ shift, $\Delta z_l^2$ & Gauss ($- 0.005$, 0.007) \\
		Lens photo-$z$ shift, $\Delta z_l^3$ & Gauss (0.006, 0.006) \\
		Lens photo-$z$ shift, $\Delta z_l^4$ & Gauss (0.000, 0.010) \\
		Lens photo-$z$ shift, $\Delta z_l^5$ & Gauss (0.000, 0.010) \\
		Source photo-$z$ shift, $\Delta z_s^1$ & Gauss ($- 0.001$, 0.016) \\
		Source photo-$z$ shift, $\Delta z_s^2$ & Gauss ($- 0.019$, 0.013) \\
		Source photo-$z$ shift, $\Delta z_s^3$ & Gauss (0.009, 0.011) \\
		Source photo-$z$ shift, $\Delta z_s^4$ & Gauss ($- 0.018$, 0.022) \\
		Multiplicative shear calibration per redshift bin (4 bins), $m_i$ & Gauss (0.012, 0.023)\\
		\hline
    \end{tabular}
     \caption{Nuisance parameter priors for each likelihood. For flat priors, the range is given. For each Gaussian prior, we report the mean and standard deviation as the pair $(\mu,\sigma)$.}
\end{table*}

\section{DES Y1 neutrino treatment}\label{app:neutrino}

\begin{figure}
\begin{center}
\includegraphics[width=0.69\textwidth]{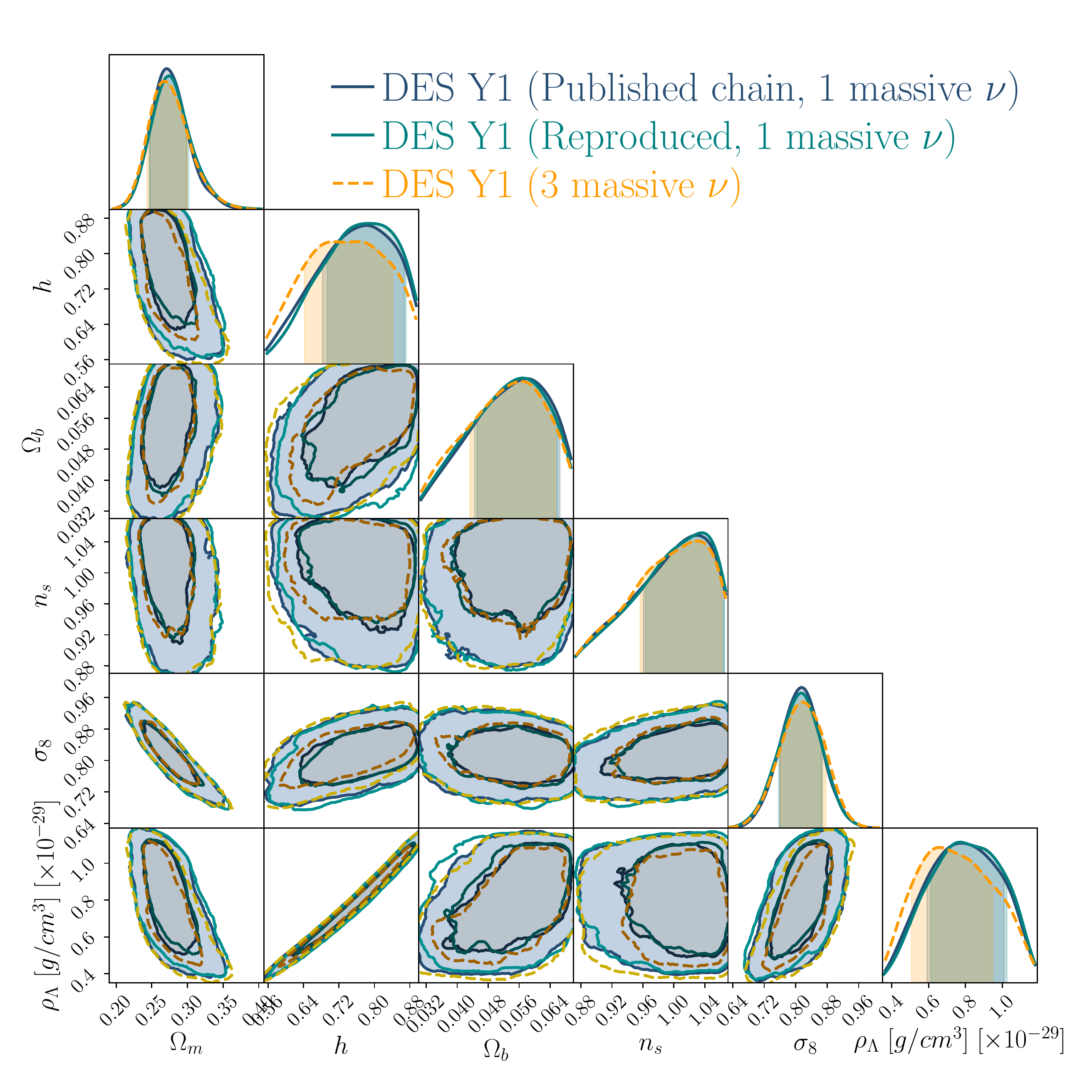}
\caption{DES Y1 3$\times$2pt likelihood comparison for different assumptions about the number of massive neutrinos. In this work we assume 3 massive neutrinos in all the likelihoods, and in this figure we check this choice does not impact the results of the DES Y1 3$\times$2pt analysis, where one neutrino species was assumed to be massive in the original publication from \cite{y1-3x2pt}. We also check we can reproduce the original results, validating our pipeline.}
\label{fig:neutrinos_y1}
\end{center}
\end{figure}

In the DES Y1 3$\times$2pt analysis \cite{y1-3x2pt}, one neutrino species was assumed to be massive. In the more recent  DES Y3 3$\times$2pt work \citep{y3-3x2pt}, three species of neutrinos were considered massive, both in the DES-only chains as well as the ones involving public data which we use in this work, such as updated CMB and Supernova measurements. We decided to also assume 3 massive neutrinos in all the likelihoods used in the current paper. In this appendix we check the impact of this choice in the DES Y1 3$\times$2pt posterior. In Fig.~\ref{fig:neutrinos_y1} we show the published and publicly available\footnote{\texttt{https://des.ncsa.illinois.edu/releases/y1a1/key-products}} DES Y1 3$\times$2pt chain, our reproduction of that chain using the public version of \textsc{Cosmosis} but still one massive neutrino, and then switching to assuming 3 massive neutrinos. The posterior differences are not significant.

\section{Examples of theoretical concepts linking  $\rho_\Lambda$ with laboratory measurements}\label{sec:proposals}

In this section we review  some examples of existing theoretical proposals that have the potential to connect the value of $\rho_\Lambda$ to other physical quantities measured in the laboratory, such as the masses of elementary particles. All of these connective ideas are still in the conceptual stage, but they illustrate the class of theory that could be developed to make a meaningful prediction for the absolute value of $\Lambda$. 

The cosmological constant $\Lambda$ of Einstein's field equations was originally conceived as an invariant parameter governing space-time curvature or geometry.  Its physical effect is to produce a relative acceleration of two particles in an empty space with zero density. The physical effect is equivalent to that of a nonzero energy-momentum tensor $T^{\mu\nu}$ for ``empty'' space: $\rho_\Lambda c^2= T^{\rm vac}_{00}$. The latter interpretation is natural in modern quantum field theory, where the physical vacuum state can have an arbitrary nonzero mean gravitating energy density $\rho_\Lambda$, often referred to as ``Dark Energy''. 
  
The actual physical value of $\rho_\Lambda$ is determined by unknown physics of quantum gravity \citep{Weinberg1989}. A  na\"ive picture based on effective quantum field theory leads to an estimated value that depends on an assumed resolution or ``UV cutoff'' scale.  In this picture, the vacuum has a gravitating energy $E=mc^2= \hbar\omega/2$ for each  mode of frequency $\omega$, corresponding to the mode's quantum zero-point fluctuation energy, but only up to some arbitrary cutoff scale:   there is assumed to be zero gravity for modes with energy exceeding  $mc^2$, or equivalently,  wavenumbers exceeding $mc/\hbar$.  Since linearized quantum fields are formally consistent with  gravitational interactions up to the Planck mass scale $m_P\equiv\sqrt{\hbar c/G}$, it is natural to write the dependence on cutoff mass parameter $m$ in Planck units,
\begin{equation}\label{extradimensionscaling}
    \rho_\Lambda\simeq (m/m_P)^4 \rho_P,
\end{equation}
where 
\begin{equation}\label{Planckdensity}
 \rho_P=  m_P^4 c^3/\hbar^3  
\end{equation}
denotes the Planck mass-energy density.

\begin{figure*}
\begin{center}
\includegraphics[width=0.80\textwidth]{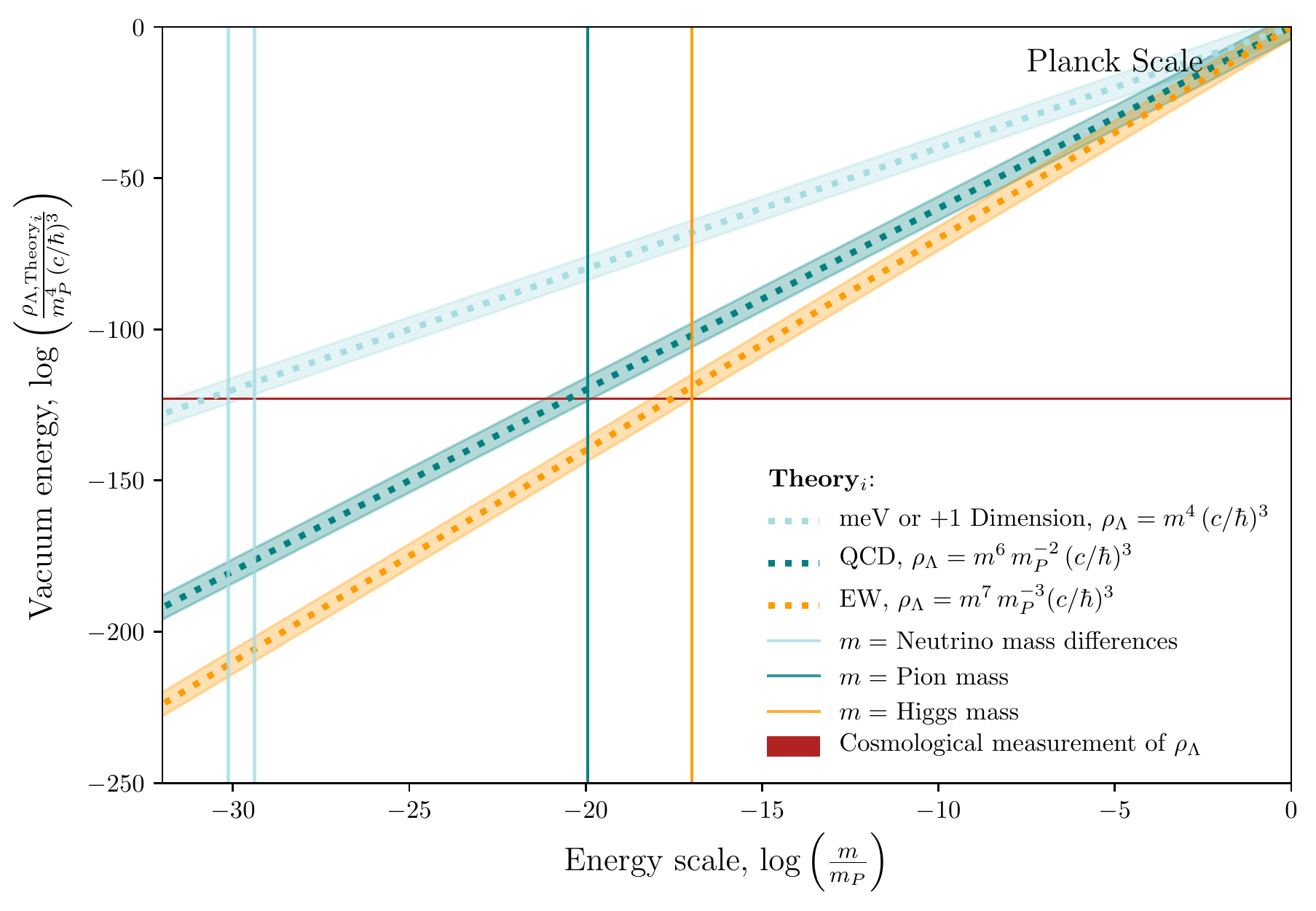}
\caption{Schematic comparison summarizing several existing theoretical  proposals to connect a laboratory-measured physical mass scale (horizontal axis) to  $\rho_\Lambda$ (vertical axis), via a small quantum-gravitational correction.  The Planck scale for mass and density is at the upper right corner. The horizontal axis shows a fundamental physical quantity that can be measured in the laboratory, such as a particle mass, or a length expressed in mass units, $\lambda=\hbar/mc$.   In each theory, the small value of the cosmological constant (shown in red) is connected with some other fixed scale of physics that is far below the Planck energy. The sloping dotted bands labelled as Theory$_i$ show the scaling predicted for how the gravitational density of the low-temperature physical vacuum $\rho_\Lambda$ depends on the physical mass scale for each of the theories described in Sec~\ref{sec:proposals}.
The width of the bands shows schematically that although the slope is predicted, the normalization is not yet precisely calculated in any of the concepts; a fully formulated theory would be a line. The power laws shown connect $\rho_\Lambda$ to measurements at the electroweak symmetry breaking scale scale (EW), the QCD scale (Eq.~\ref{QCDscaling}), or  modifications of gravity  at  milli-eV  scale  ($mc^2\sim$ meV in Eq.~\ref{extradimensionscaling}), which may connect to direct gravity experiments  on the submillimeter scale, or to neutrino masses. Vertical lines show examples of particle mass measurements \citep{PDG2020}; the masses shown for neutrinos, 0.05 eV and 0.009 eV, are respective values from atmospheric and solar oscillation squared-mass differences, $\sqrt{\Delta m^2}$. A precisely formulated theory  normalized by laboratory measurements will be tested by comparison with absolute measurements of $\rho_\Lambda$, as implemented here. The thin red horizontal band shows the fiducial measurement of $\rho_\Lambda$ we present in this work together with its uncertainty, which is too small to appreciate in the plot. 
%The measured $\rho_\Lambda$ today corresponds to lower vacuum energies in all the theories. $\rho_\Lambda$ is a constant in our universe within the $\Lambda$CDM model, but during inflation it could have had a different much larger value, which is predicted by the theories displayed above. 
%Schematic comparison of theory concepts with data in Planck units.  %Horizontal axis shows the energy scale $E$ that can be measured in the laboratory, vertical axis shows the gravitational density of the low-temperature physical vacuum $\rho_\Lambda$ in various theories. The lines show proposed power laws for theories based on the electroweak scale, the QCD scale, or  modifications of gravity by an extra dimension at the submillimeter scale. Theories normalized by   laboratory measurements will be tested by comparison with absolute measurements of  $\rho_\Lambda$. 
}
\label{scalecomparison}
\end{center}
\end{figure*}

If this estimate of  field-vacuum gravitation were valid up to the  Planck scale $m \sim  m_P$, the predicted  density would be about 122 orders of magnitude larger than observed (Fig.~\ref{scalecomparison}), which is famously known as the cosmological constant problem. Clearly, this picture must break down in a fundamental way at some scale $m$ far below $m_P$. Apparently, some new set of rules, beyond those included in quantum field theory alone,  governs the relationship of gravity (or geometry) with quantum matter. 

One possibility  is that a fundamental symmetry of quantum gravity guarantees a near vanishing of $\rho_\Lambda$, with  a  nonzero correction that is very small in Planck units \citep{Weinberg1989}. However, there is no general agreement about the origin of either the symmetry or the correction, which is one reason for the widely held view that the value of $\rho_\Lambda$ is set by chance and has no particular significance.  If such a correction exists, it could be calculable in a correct theory of quantum gravity.

\begin{comment}
\begin{equation}
\rho =  \frac{m}{V} = \frac{m}{\lambda^3}
\end{equation}

Planck length:
\begin{equation}
l_P =  \sqrt{\frac{\hbar G}{c^3}}
\end{equation}
Planck mass:
\begin{equation}
m_P =  \sqrt{\frac{\hbar c}{G}}
\end{equation}

\begin{equation}
\rho_P =  m_P^4 c^3/\hbar^3   = \frac{ c^5}{\hbar G^2}
\end{equation}
\end{comment}

Although no well developed theory for the absolute value of $\rho_\Lambda$ currently exists, it is well known that consistency with gravity adds new constraints to quantum states of fields in the infrared \cite{CohenKaplanNelson1999, cohen2021}, which are not included in the linearized quantum gravity theory that leads to the wrong answer for $\rho_\Lambda$.  Thus, there is the possibility that a deeper theory of quantum gravity will connect the  small  nonzero value for the gravitation of the physical vacuum with matter interactions that can be measured in the laboratory. Such a theory would  precisely connect the large dimensionless number $\rho_P/\rho_\Lambda$ with other large dimensionless numbers in physics. 

For example, the density of zero-point  field fluctuations with a cutoff (Eq.~\ref{extradimensionscaling}) agrees with the observed value  of $\rho_\Lambda$ for an energy cut-off scale $mc^2$ in the milli-eV  range,  so that their gravity is suppressed for  wavelengths smaller than $\hbar/mc\simeq 100\mu m$. It has been proposed that  extra compactified dimensions could lead to  modification of the direct gravitational force below the compactification length scale. Many  implementations of this idea  have been ruled out by laboratory experiments \cite{Kapner2007, adelberger2009}, but some versions  may still be viable \citep{Dupays2013}. In a string landscape, the scale of an extra dimension could manifest in measured neutrino masses on a similar scale \citep{dey2009,Ibanez2017}. 

A related proposal also  invoking the milli-eV  scale posits that the fundamental vacuum energy of the universe is in fact zero (e.g., due to a symmetry of the field ground state); if that is the case, then the effective vacuum energy density at any epoch will be dominated by the heaviest field that has not yet reached its ground state. At late times, such a field must be very light. Some work has been done exploring a simple chiral symmetry breaking model in which this energy density is of order $\sum_i m_{\nu,i}^4$, in the correct range for neutrino masses $m_{\nu}$ in the milli-eV range \cite{Frieman1995} . In this case, a precision measurement of $\rho_\Lambda$ would constrain a sum of powers of light neutrino masses.

Other conceptual proposals connect $\rho_\Lambda$ directly with  scales of Standard Model particles and fields. For example, it has been suggested that a small non-zero cosmological constant might be caused by gravitational quantum corrections from QED or nuclear interactions \cite{zeldovich1967, Zeldovich1968}. Although this does not occur in linearized quantum gravity, it might arise from causally-coherent quantum gravity coupled to nonlocal virtual states of a strongly self-interacting gluonic field vacuum \citep{Hogan2020}. Space-time averages of correlated field fluctuations encompass roughly the space-time volume of a virtual pion, so it is natural to use the pion mass $m_\pi$ to normalize to the physical QCD scale:
\begin{equation}\label{QCDscaling}
    \rho_\Lambda  = \Lambda_\pi\  ({m_\pi/ m_P})^6\  \rho_P,
\end{equation} 
where $\Lambda_\pi$ is a dimensionless coefficient. According to this conjecture, the value of $\Lambda$ is determined by the gravity of virtual states of the same QCD fields that determine the measured pion mass $m_\pi$, so a direct calculation of $\Lambda_\pi$ may be possible from already-known equations of QCD\footnote{This scenario accounts in broad terms  for the famous ``Why Now?'' coincidence associated with the value of $\rho_\Lambda$, the fact that  the fundamental acceleration timescale determined by $\rho_\Lambda$ is of the same order of magnitude as the current age of the universe when we happen to be viewing it.
In this scenario, the  timescale for cosmic acceleration, $(m_P/m_\pi)^3 t_P$, depends on QCD in the the same  way as the large dimensionless number $(m_P/m_{\text{proton}})^3 t_P$ that governs the astrophysical evolution of stellar systems, as well as  biological systems that depend on the elements and energy they generate. The large ratio of scales itself is widely attributed to the logarithmic running of QCD coupling in the context of a unified field theory, so both numbers have the same exponential dependence on the strong coupling constant \citep{Hogan2000}.}

Similarly,  it has been  suggested that  emergent gravity in a discrete system  connected to  electroweak  interactions leads to  $\rho_\Lambda$ in Planck units  roughly equal to the seventh power of the Higgs mass \cite{Perez2018, Perez2019}.
These conceptual proposals are illustrated schematically  in Fig.~\ref{scalecomparison}.

A well-developed predictive theory for the absolute value of the cosmological constant   will require a deeper theory of the relationship of the quantum states of gravity with those of particle fields. An accurate comparison with $\rho_\Lambda$ would provide a fundamental test of such a theory, in much the same way that  an accurate measurement of the electron magnetic moment tests  quantum electrodynamics. Any physical theory that connects the cosmological constant to  laboratory-measured quantities should predict a value for $\rho_\Lambda$ in absolute physical units.

To implement a comparison of this kind, the absolute physical value of the vacuum energy density needs to be measured in a way that allows its value to be compared as directly as possible with laboratory measurements of physical constants, including systematic errors and covariances. Ideally, the cosmological measurement should aim to achieve an accuracy comparable to the laboratory measurements it will be compared with. Because $\rho_\Lambda$ depends on a high power of locally measured scales of mass or energy, a realistic  cosmological precision of a few percent is equivalent to sub-percent precision in the lab. 

Currently,  no  theory is developed well enough to allow for such a precise experimental test: the normalization in the conceptual proposals just cited is only worked out to order-of-magnitude precision. Thus at present, the main limit is the quality of theoretical predictions. Still, as we show here, the data are already available for a meaningful test, should  a precisely developed theory be developed. The accuracy of measurements should set the aspirational precision of theories.

\acknowledgments
  This work was supported by the Fermi National Accelerator Laboratory (Fermilab), a U.S. Department of Energy, Office of Science, HEP User Facility, managed by Fermi Research Alliance, LLC (FRA), acting under Contract No. DE-AC02-07CH11359. JP and CC are supported by DOE grant DE-SC0021949. We thank Joe Zuntz for his support on \textsc{CosmoSIS}.
  
  This project used public data based on observations obtained with Planck (http://www.esa.int/Planck), an ESA science mission with instruments and contributions directly funded by ESA Member States, NASA, and Canada.
 
  This project used public archival data from the Dark Energy Survey (DES). Funding for the DES Projects has been provided by the U.S. Department of Energy, the U.S. National Science Foundation, the Ministry of Science and Education of Spain, the Science and Technology FacilitiesCouncil of the United Kingdom, the Higher Education Funding Council for England, the National Center for Supercomputing Applications at the University of Illinois at Urbana-Champaign, the Kavli Institute of Cosmological Physics at the University of Chicago, the Center for Cosmology and Astro-Particle Physics at the Ohio State University, the Mitchell Institute for Fundamental Physics and Astronomy at Texas A\&M University, Financiadora de Estudos e Projetos, Funda{\c c}{\~a}o Carlos Chagas Filho de Amparo {\`a} Pesquisa do Estado do Rio de Janeiro, Conselho Nacional de Desenvolvimento Cient{\'i}fico e Tecnol{\'o}gico and the Minist{\'e}rio da Ci{\^e}ncia, Tecnologia e Inova{\c c}{\~a}o, the Deutsche Forschungsgemeinschaft, and the Collaborating Institutions in the Dark Energy Survey. The Collaborating Institutions are Argonne National Laboratory, the University of California at Santa Cruz, the University of Cambridge, Centro de Investigaciones Energ{\'e}ticas, Medioambientales y Tecnol{\'o}gicas-Madrid, the University of Chicago, University College London, the DES-Brazil Consortium, the University of Edinburgh, the Eidgen{\"o}ssische Technische Hochschule (ETH) Z{\"u}rich,  Fermi National Accelerator Laboratory, the University of Illinois at Urbana-Champaign, the Institut de Ci{\`e}ncies de l'Espai (IEEC/CSIC), the Institut de F{\'i}sica d'Altes Energies, Lawrence Berkeley National Laboratory, the Ludwig-Maximilians Universit{\"a}t M{\"u}nchen and the associated Excellence Cluster Universe, the University of Michigan, the National Optical Astronomy Observatory, the University of Nottingham, The Ohio State University, the OzDES Membership Consortium, the University of Pennsylvania, the University of Portsmouth, SLAC National Accelerator Laboratory, Stanford University, the University of Sussex, and Texas A\&M University. Based in part on observations at Cerro Tololo Inter-American Observatory, National Optical Astronomy Observatory, which is operated by the Association of Universities for Research in Astronomy (AURA) under a cooperative agreement with the National Science Foundation.

Funding for the Sloan Digital Sky 
Survey IV has been provided by the 
Alfred P. Sloan Foundation, the U.S. 
Department of Energy Office of 
Science, and the Participating 
Institutions. SDSS-IV acknowledges support and 
resources from the Center for High 
Performance Computing  at the 
University of Utah. The SDSS 
website is www.sdss.org. SDSS-IV is managed by the 
Astrophysical Research Consortium 
for the Participating Institutions 
of the SDSS Collaboration including 
the Brazilian Participation Group, 
the Carnegie Institution for Science, 
Carnegie Mellon University, Center for 
Astrophysics | Harvard \& 
Smithsonian, the Chilean Participation 
Group, the French Participation Group, 
Instituto de Astrof\'isica de 
Canarias, The Johns Hopkins 
University, Kavli Institute for the 
Physics and Mathematics of the 
Universe (IPMU) / University of 
Tokyo, the Korean Participation Group, 
Lawrence Berkeley National Laboratory, 
Leibniz Institut f\"ur Astrophysik 
Potsdam (AIP),  Max-Planck-Institut 
f\"ur Astronomie (MPIA Heidelberg), 
Max-Planck-Institut f\"ur 
Astrophysik (MPA Garching), 
Max-Planck-Institut f\"ur 
Extraterrestrische Physik (MPE), 
National Astronomical Observatories of 
China, New Mexico State University, 
New York University, University of 
Notre Dame, Observat\'ario 
Nacional / MCTI, The Ohio State 
University, Pennsylvania State 
University, Shanghai 
Astronomical Observatory, United 
Kingdom Participation Group, 
Universidad Nacional Aut\'onoma 
de M\'exico, University of Arizona, 
University of Colorado Boulder, 
University of Oxford, University of 
Portsmouth, University of Utah, 
University of Virginia, University 
of Washington, University of 
Wisconsin, Vanderbilt University, 
and Yale University.

\bibliographystyle{ieeetr}
\bibliography{library}


%\begin{thebibliography}{99}

%\bibitem{a}
%Author, \emph{Title}, \emph{J. Abbrev.} {\bf vol} (year) pg.

%\bibitem{b}
%Author, \emph{Title},
%arxiv:1234.5678.

%\bibitem{c}
%Author, \emph{Title},
%Publisher (year).

% Please avoid comments such as "For a review'', "For some examples",
% "and references therein" or move them in the text. In general,
% please leave only references in the bibliography and move all
% accessory text in footnotes.

% Also, please have only one work for each \bibitem.

%\end{thebibliography}
\end{document}